\documentclass[prd,twocolumn,unsortedaddress,superscriptaddress]{revtex4-2}
\usepackage[english]{babel}
\addto\captionsenglish{}
\usepackage{graphicx}

\usepackage{epsfig,color}
\usepackage{amsmath,amssymb,calc}
\usepackage{amsthm}
\usepackage{enumerate}
\usepackage{hyperref}
\usepackage{amsxtra,amsfonts,bm,txfonts}
\usepackage{blkarray}
\usepackage{adjustbox}
\usepackage{braket}
\usepackage{adjustbox}
\usepackage{xcolor}
\usepackage{physics}

\begin{document}

\title{Fermionic Gaussian Projected Entangled Pair States in \texorpdfstring{$3+1d$}{3+1d}: Rotations and Relativistic Limits}

\date{\today}

\author{Patrick Emonts}
\affiliation{Lorentz Institute, Leiden University, Niels Bohrweg 1, 2333 CA Leiden, Netherlands}
\author{Erez Zohar}
\affiliation{Racah Institute of Physics, The Hebrew University of Jerusalem, Givat Ram, Jerusalem 91904, Israel}

\begin{abstract}
Fermionic Gaussian Projected Entangled Pair States (PEPS) are fermionic tensor network state constructions which describe the physics of ground states of non-interacting fermionic Hamiltonians. 
As non-interacting states, one may study and analyze them very efficiently, in both analytical and numerical means. 
Recently it was shown that they may be used as the starting point - after applying so-called PEPS gauging mechanisms - for variational study of non-trivial, interacting states of lattice gauge theories. This is done using sign-problem free variational Monte-Carlo. 
In this work we show how to generalize such states from two to three spatial dimensions, focusing on spin representations and requirements of lattice rotations. 
We present  constructions which are crucial for the application of the above mentioned variational Monte-Carlo techniques for studying non-perturbative lattice gauge theory physics, with fermionic matter, in $2+1$-d and $3+1$-d models. 
Thus, the constructions presented here are crucial for the study of non-trivial lattice gauge theories with fermionic tensor network states.
\end{abstract}

\maketitle

\section{Introduction}

Tensor network states (TNS), and in particular Matrix Product States (MPS) and Projected Entangled Pair States (PEPS)~\cite{orus_practical_2014,cirac_renormalization_2009,cirac_matrix_2020,white_density_1992,schollwock_density-matrix_2011} have been very successful as ansatz states for the study of complex, strongly-correlated quantum many body systems  thanks to their built-in entanglement entropy area law~\cite{eisert_area_2010}. 

Among the many members of the TNS family, an important one is that of fermionic TNS, and in particular fermionic PEPS~\cite{corboz_simulation_2010,kraus_fermionic_2010}. 
With their special statistics, fermions require special treatment and cautious analysis while being used as building blocks for tensor network states, just like in any other physical instance; and since fermions are so fundamental both in condensed matter and in particle physics, and appear in a variety of strongly correlated models, studying them using entanglement based approaches such as tensor network states is a very promising research avenue.

Gaussian fermionic PEPS~\cite{kraus_fermionic_2010} are a particular type of fermionic PEPS, in which the PEPS formalism is used for constructing \emph{fermionic Gaussian states}~\cite{bravyi_lagrangian_2005,surace_fermionic_2022} - ground states of free, non-interacting fermionic Hamiltonians. 
Such states are fully characterized using their covariance matrices~\cite{bravyi_lagrangian_2005}, which are in the core of the \emph{Gaussian formalism} that enables very efficient computations. 
While one may argue that free fermionic states are trivial, this is not the case. 
Gaussian fermionic PEPS have been used, for example, for analyzing models with chiral topological order~\cite{wahl_projected_2013,wahl_symmetries_2014,yang_chiral_2015} and for showing that tensor networks can resolve Fermi surfaces~\cite{mortier_tensor_2022}.

In the context of particle physics, a lot of research has been made in the last decade, generalizing and applying the TNS framework for the study of lattice gauge theories (see, e.g., the review papers~\cite{dalmonte_lattice_2016,banuls_review_2020,montangero_loop_2022}, as well as~\cite{meurice_tensor_2022} on a complementary, non-Hamiltonian approach, and references therein). 
In such theories~\cite{wilson_confinement_1974,kogut_hamiltonian_1975,kogut_introduction_1979} the matter is mostly fermionic. 
In minimally coupled lattice gauge theory Hamiltonians, all the terms involving fermionic operators are quadratic in the fermions, which allows one to use, as a PEPS ansatz, Gauged Gaussian fermionic PEPS~\cite{zohar_combining_2018}. 
There, thanks to the local (gauge) symmetry, even though the full state is interacting and non-Gaussian, one can use an expansion in Gaussian states for efficiently computing, using the Gaussian formalism, relevant observables using a sign-problem free Monte-Carlo integration. 
This technique, while being based on a tensor network construction, which brings along advantages such as the entanglement area-law and easy encoding of symmetries, does not rely on conventional tensor network contraction methods, and thus is free of the problems they involve in high spatial dimensions. 
It is  based on the gauge symmetry and the properties of the gauging procedure and minimal coupling, and  was benchmarked for pure gauge models~\cite{emonts_variational_2020,emonts_finding_2023}.

Previous analytical studies and benchmarks of this method have only focused on $2+1-d$ models~\cite{zohar_fermionic_2015,zohar_projected_2016}. 
However, using the gauging procedure introduced in~\cite{zohar_building_2016}, it is possible to obtain the relevant ansatz states once a matter-only PEPS which is globally invariant under the gauge group is built, in any spatial dimension~\cite{zohar_combining_2018}. 
To extend the computation method to higher dimensions, it is important to generalize the existing constructions of relativistic fermionic Gaussian PEPS in $2+1$ dimensions to $3+1$ dimensions too. 
In this work we consider this problem, and show how to construct Gaussian fermionic PEPS in $d=2,3$ space dimensions with the right lattice rotation invariance properties which are crucial for obtaining reasonable, relativistic continuum limits. 
While this was previously done in $d=2$~\cite{zohar_fermionic_2015}, here we put the derivation in a more physical context and extend it to $d=3$.

This is a constructive, analytical work, aimed at introducing the missing element to be used in applying the methods of~\cite{zohar_combining_2018} for lattice gauge theories in three spatial dimensions. 
As such, it involves no numerical results. 
Since it studies the free states to be later gauged using \cite{zohar_building_2016}, everything can be done completely analytically.

We will begin by discussing rotation properties of Gaussian fermionic states in $d=2,3$ (section~\ref{sec:free_fermionic_states}) demonstrate that the fermions follow, in a sense, relativistic rotation properties, even in scenarios which are sometimes referred to as "spinless". 
In section~\ref{sec:peps_construction}, we will show how to construct Gaussian PEPS with these properties in various ways that account for spin (corresponding to various prescriptions for dealing with the fermion doubling issue~\cite{nielsen_no-go_1981}) and conclude in section~\ref{sec:free_fermionic_states} by providing exact ground state constructions of some relativistic free fermionic models using Gaussian fermionic PEPS. 
As this paper is aimed at people from both the tensor network and the lattice gauge theory communities, and since these states, even after gauging, will not be studied using tensor network contracting schemes, we have not made use of tensor network digramamatic language here.

\section{Free Fermionic States}\label{sec:free_fermionic_states}

Before considering PEPS in particular, let us focus on the lattice rotation properties of a wider set of states - fermionic Gaussian states.

We consider a square / cubic lattice in $d$ spatial dimensions, whose sites are labeled by vectors of integers, $\mathbf{x}=\left(x_1,...,x_d\right) \in \mathbb{Z}^d$. 
On each site $\mathbf{x}$ introduce a set of fermionic modes, created by the operators $\psi_a^{\dagger}\left(\mathbf{x}\right)$ and annihilated by $\psi_a\left(\mathbf{x}\right)$. 
$a$ may be some collection of indices corresponding to quantum numbers such as spin, flavor, color etc., depending on the physical context. 
These fermionic Fock operators satisfy the usual anti-commutation relations
\begin{equation}
\left\{\psi_a\left(\mathbf{x}\right),\psi_b\left(\mathbf{y}\right)\right\}=\left\{\psi_a^{\dagger}\left(\mathbf{x}\right),\psi_b^{\dagger}\left(\mathbf{y}\right)\right\}=0
\label{eqacom1}
\end{equation}
and
\begin{equation}
	\left\{\psi_a\left(\mathbf{x}\right),\psi_b^{\dagger}\left(\mathbf{y}\right)\right\}=\delta_{\mathbf{x},\mathbf{y}}.
	\label{eqacom3}
\end{equation}

Free fermionic states are ground states of free fermionic Hamiltonians, which are quadratic in the fermionic operators. 
There always exists a canonical transformation allowing to convert any such Hamiltonian to a superconducting form, for which the ground state will be a BCS (Bardeen–Cooper–Schrieffer) state. 
This means that we are allowed, without losing any generality, to express any fermionic Gaussian state in the form
\begin{equation}
	\left|\psi\right\rangle = \exp(\underset{\mathbf{x},\mathbf{y}}{\sum}T_{ab}\left(\mathbf{x},\mathbf{y}\right)\psi_a^{\dagger}\left(\mathbf{x}\right)\psi_b^{\dagger}\left(\mathbf{y}\right))\ket{\Omega},
	\label{psidef}
\end{equation}
where $\ket{\Omega}$ is the Fock vacuum, annihilated by all the $\psi_a\left(\mathbf{x}\right)$ operators,
\begin{equation}
\psi_a\left(\mathbf{x}\right)\ket{\Omega} = 0, \quad \forall \mathbf{x}\in \mathbb{Z}^d,a.
\end{equation}

All the physical information is stored in the tensor $T_{ab}\left(\mathbf{x},\mathbf{y}\right)$. 
Due to the fermionic anti-commutation relation~\eqref{eqacom1}, it must be anti-symmetric,
\begin{equation}
T_{ba}\left(\mathbf{y},\mathbf{x}\right) = -T_{ab}\left(\mathbf{x},\mathbf{y}\right).
\label{eqasym}
\end{equation}
Further properties are dictated by the physical features of the state, including its symmetries, and in particular the rotations. 
Let us consider the rotation of such fermionic operators for spatial dimensions $d=2,3$.

\subsection{\texorpdfstring{$d=2$}{d=2} Rotations}
 In $d=2$, we introduce the $\pi/2$ rotation operation of the coordinates by
\begin{equation}
	\Lambda\mathbf{x}=\left(-x_2,x_1\right).
\end{equation}
As expected, $\Lambda^4 \mathbf{x} = \mathbf{x}$.

The rotation of the fermionic creation operators is carried out by a unitary operator $\mathcal{U}_R$. 
In order to study the effect of rotation while describing it in the simplest possible way, let us separate between the indices within $a$ which are affected by rotations (spin) and those which do not (color, flavor, etc.). 
Since we only care about rotations at the moment, we shall omit the non-spin indices, and leave $a$ exclusively for the spin. 
Then, we can simply implement the rotation by
\begin{equation}
\mathcal{U}_R\psi^{\dagger}_a\left(\mathbf{x}\right)\mathcal{U}^{\dagger}_R = \eta_{ab}\left(\mathbf{x}\right)\psi^{\dagger}_b\left(\Lambda\mathbf{ x}\right)
\label{rot2}
\end{equation}
where $\eta\left(\mathbf{x}\right)$ is some unitary matrix. 
Invariance of the state $\left|\psi\right\rangle$ under four rotations, even in the non rotational-invariant case, implies that for every $\mathbf{x}$,
\begin{equation}
	\eta\left(\mathbf{x}\right)\eta\left(\Lambda\mathbf{x}\right)\eta\left(\Lambda^2\mathbf{x}\right)\eta\left(\Lambda^3\mathbf{x}\right) = \pm\mathbf{1}.
	\label{eqprodpm1}
\end{equation} 
If we further wish to assume that the state is invariant under rotations, we also obtain that 
\begin{equation}
\eta_{ac}\left(\mathbf{x}\right)T_{cd}\left(\mathbf{x},\mathbf{y}\right)\eta^T_{db}\left(\mathbf{y}\right) = T_{ab}\left(\Lambda\mathbf{x},\Lambda\mathbf{y}\right),
\label{eqrot}
\end{equation}
for every $\mathbf{x},\mathbf{y}$.

Suppose we wish to consider the simplest case, in which we have a single spin component, making $\eta$ simply a phase. 
Furthermore, we wish to see if it is possible to consider full translational invariance, both for $T$ and for $\eta$. 
In such a case, Eq. \eqref{eqprodpm1} significantly simplifies to 
\begin{equation}
\eta^4=\pm1
\end{equation}
In this case, consider the double rotation ($\Lambda^2\mathbf{x}=-\mathbf{x}$), and focus on $T\left(\mathbf{x},\mathbf{x}+\mathbf{v}\right)$, for some fixed  vector $\mathbf{v}$. 
If we have rotation invariance, Eq.~\eqref{eqrot} implies that
\begin{equation}
	T\left(-\mathbf{x},-\mathbf{x}-
	\mathbf{v}\right)=\eta^4 T\left(\mathbf{x},\mathbf{x}+\mathbf{v}\right).
\end{equation}
Eq.~\eqref{eqasym} tells us that due to the fermionic anti-commutation~\eqref{eqacom1},
\begin{equation}
	T\left(-\mathbf{x},-\mathbf{x}-
	\mathbf{v}\right)=-T\left(-\mathbf{x}-
	\mathbf{v},-\mathbf{x}\right),
\end{equation}
and translation invariance implies that 
\begin{equation}
		T\left(-\mathbf{x}-
		\mathbf{v},-\mathbf{x}\right)=T\left(\mathbf{x},\mathbf{x}+\mathbf{v}\right).
\end{equation}
From all this we conclude that if we want both single-site translation invariance and rotation invariance of the state $\left|\psi\right\rangle$ from Eq.~\eqref{psidef} in $d=2$, we must have
\begin{equation}
	\eta^4=-1
\end{equation}
which is a manifestation of the spin-statistics theorem in this $d=2$ "spinless" case with translation and rotation invariance: a complete rotation of a fermion still gives rise to a sign, even though we do not need any actual spin components. This implies that this general set of states could describe some relativistic physics in the continuum limit, in a sense, because it has the right symmetry properties.

To give it a more concrete physical meaning, let us recall Susskind's staggered fermions~\cite{susskind_lattice_1977}. In that model, the components of a continuum spinor are distributed across lattice sites, giving rise to a lattice model of "spinless" fermions, whose continuum limit has a reduced amount of fermionic doubling~\cite{nielsen_no-go_1981}. Let us see, then, that the staggered fermions Hamiltonian in $d=2$ has the same symmetry properties.

The original Hamiltonian~\cite{susskind_lattice_1977} takes the form
\begin{widetext}
	\begin{equation}
		H=M\underset{\mathbf{x}}{\sum}\left(-1\right)^{x_1+x_2}\psi^{\dagger}\left(\mathbf{x}\right)\psi\left(\mathbf{x}\right) + 
		\left(
		\frac{i}{2a} \underset{\mathbf{x}}{\sum} \psi^{\dagger}\left(\mathbf{x}\right)\psi\left(\mathbf{x}+\hat{\mathbf{e}_1}\right)
		-\frac{1}{2a}\left(-1\right)^{x_1+x_2} \underset{\mathbf{x}}{\sum} \psi^{\dagger}\left(\mathbf{x}\right)\psi\left(\mathbf{x}+\hat{\mathbf{e}_2}\right)
		+\text{h.c.}
		\right)
	\end{equation}
\end{widetext}
where $\hat{\mathbf{e}}_i$ are lattice vectors and $a$ is the lattice spacing. Here, fermions on even sites represent particles, while the absence thereof on odd sites represents antiparticles. In a careful continuum limit, these will unite to a two component spinor, as required in a $2+1$ dimensional Dirac theory (without chirality). Given this insight, it is not a surprise that in a $\pi/2$ rotation, the creation operators on the even sublattice are transformed with a phase $\eta=e^{i\pi/4}$, and those on the odd sublattice with $\overline{\eta}$ (note that the Hamlitonian is invariant under such rotations); the only relevant rotation would be implemented, in the continuum, by $e^{i\pi\sigma_3/4}$, a diagonal matrix with these two phases, to be associated with the two sublattices when staggered on the lattice.

If we perform a particle-hole transformation on the odd sublattice, that is
\begin{equation}
	\psi^{\dagger}\left(\mathbf{x}\right) \leftrightarrow \psi\left(\mathbf{x}\right),\quad \forall \mathbf{x}:x_1+x_2\quad \text{odd}.
\end{equation}
Then, up to an irrelevant constant,
\begin{widetext}
	\begin{equation}
		H=M\underset{\mathbf{x}}{\sum}\psi^{\dagger}\left(\mathbf{x}\right)\psi\left(\mathbf{x}\right) + 
		\left(
		\frac{i}{2a} \underset{\mathbf{x}}{\sum} \psi^{\dagger}\left(\mathbf{x}\right)\psi^{\dagger}\left(\mathbf{x}+\hat{\mathbf{e}}_1\right)
		-\frac{1}{2a} \underset{\mathbf{x}}{\sum} \psi^{\dagger}\left(\mathbf{x}\right)\psi^{\dagger}\left(\mathbf{x}+\hat{\mathbf{e}}_2\right)
		+\text{h.c.}
		\right)
		\label{suss2d}
	\end{equation}

\end{widetext}
Note that this model is no longer staggered (has a single site translation invariance), and it is rotationally invariant with a fixed $\eta=e^{i\pi/4}$ whose fourth power is $-1$. It is a free model whose ground state can be shown to  take the form of  $\left|\psi\right\rangle$ of Eq.~\eqref{psidef}.

\subsection{\texorpdfstring{$d=3$}{d=3} Rotations}
In $d=3$ we consider three different $\pi/2$ rotations
\begin{equation}
\begin{aligned}
	\Lambda_1\mathbf{x}=\left(x_1,-x_3,x_2\right), \\
	\Lambda_2\mathbf{x}=\left(x_3,x_2,-x_1\right),\\
	\Lambda_3\mathbf{x}=\left(-x_2,x_1,x_3\right).
\end{aligned}
\end{equation}
It is easy to see that the relation among these operations, being a subgroup of the $SO(3)$ group of continuous rotations in $d=3$, satisfied the expected anti-commutation properties, relating all three rotations, through
\begin{equation}
	\Lambda_2 = \Lambda_1^{-1}\Lambda_3\Lambda_1
	\label{alg}
\end{equation}
and similar relations.

For each rotation axis $i=1,2,3$, we introduce a unitary rotation operator $\mathcal{U}_R^{(i)}$, and unitary matrices acting on the spin indices, $\eta^{(i)}\left(\mathbf{x}\right)$, such that
\begin{equation}
	\mathcal{U}^{(i)}_R\psi^{\dagger}_a\left(\mathbf{x}\right)\mathcal{U}^{(i)\dagger}_R = \eta^{(i)}_{ab}\left(\mathbf{x}\right)\psi^{\dagger}_b\left(\Lambda\mathbf{ x}\right)
	\label{rot3}
\end{equation}
[generalizing Eq.~\eqref{rot2}]. 
Another immediate generalization from the $d=2$ case implies that
for every $\mathbf{x}$ and $i=1,2,3$,
\begin{equation}
	\eta^{(i)}\left(\mathbf{x}\right)\eta^{(i)}\left(\Lambda_i\mathbf{x}\right)\eta^{(i)}\left(\Lambda_i^2\mathbf{x}\right)\eta^{(i)}\left(\Lambda_i^3\mathbf{x}\right) = \pm\mathbf{1}.
	\label{eqprodpm13}
\end{equation} 
(invariance of a complete rotation). Again,
if we further wish to assume that the state is invariant under rotations, we obtain that 
\begin{equation}
	\eta^{(i)}_{ac}\left(\mathbf{x}\right)T_{cd}\left(\mathbf{x},\mathbf{y}\right)\eta^{(i)T}_{db}\left(\mathbf{y}\right) = T_{ab}\left(\Lambda_i\mathbf{x},\Lambda_i\mathbf{y}\right),
	\label{eqrot3}
\end{equation}
for every $\mathbf{x},\mathbf{y}$ and $i$.

\subsubsection{The "spinless", staggered case}
As for $d=2$, we begin with the "spinless" case in which we have single components on all the sites, and $\eta^{(i)}$ are simply phases. If we assume single site translation invariance, we will also get that 
\begin{equation}
\left(\eta^{(i)}\right)^4=-1
\end{equation}
 in the rotationally invariant case, for every $i$. The relations among the rotations, such as Eq.~\eqref{alg}, will tell us that all the phases are the same, $\eta^{(i)}=\eta$; the coordinate transformation (which is nothing but the spatial part of the rotation) will take care of the non-commuting nature of the rotations.

So far, everything seems nice, simple and working, but this is not the case. To see why, let us consider a simple subset of the $T$ elements, those with $\mathbf{y}=\mathbf{x}+\hat{\mathbf{e}}_i$. Translation invariance implies that
\begin{equation}
	T\left(\mathbf{x},\mathbf{x}+\hat{\mathbf{e}}_i\right) \equiv t_i
\end{equation}
independently of $\mathbf{x}$.
Rotation around the $z$ axis implies that $t_2=\eta^2 t_1$, but also that $t_3=\eta^2 t_3$. This contradicts $\eta^4=-1$, and implies that if we insist on having one spin component per site, single site translation invariance could not work (even if we set all $t_i=0$, we will run into similar problems when considering the $T$ elements between farther lattice points). This is a consequence of having a third spatial dimension; in $d=2$ we could live within the nice convenience of "spinless" fermions - they were not even staggered in the BCS picture, thanks to the fact that all rotations commute. But in $d=3$ the story is completely different.

Following what we did in the $d=2$ case, we consider Susskind's staggered fermions Hamiltonian for $d=3$~\cite{susskind_lattice_1977} and perform a particle-hole transformation on the odd sublattice, to obtain (up to an irrelevant constant)
\begin{widetext}
	\begin{equation}
		H=M\underset{\mathbf{x}}{\sum}\psi^{\dagger}\left(\mathbf{x}\right)\psi\left(\mathbf{x}\right) + 
		\left(
		\frac{i}{2a} \underset{\mathbf{x}}{\sum} \psi^{\dagger}\left(\mathbf{x}\right)\psi^{\dagger}\left(\mathbf{x}+\hat{\mathbf{e}_1}\right)
		-\frac{1}{2a} \underset{\mathbf{x}}{\sum}\left(-1\right)^{x_3} \psi^{\dagger}\left(\mathbf{x}\right)\psi^{\dagger}\left(\mathbf{x}+\hat{\mathbf{e}_2}\right)+
		\frac{i}{2a} \underset{\mathbf{x}}{\sum}\left(-1\right)^{x_1+x_2} \psi^{\dagger}\left(\mathbf{x}\right)\psi^{\dagger}\left(\mathbf{x}+\hat{\mathbf{e}_3}\right)
		+\text{h.c.}
		\right).
	\end{equation}
\label{hsussph}
This Hamiltonian is symmetric under the \emph{staggerred} (and non-commuting) rotation transformations defined with the phases
\begin{equation}
	\begin{aligned}
\eta^{(1)}\left(\mathbf{x}\right) &= \frac{1 - i\left(-1\right)^{x_1}}{\sqrt{2}}\\
\eta^{(2)}\left(\mathbf{x}\right) &= -\frac{\left(1 - i\left(-1\right)^{x_1}\right)\left(1 + i\left(-1\right)^{x_2}\right)\left(1 + i\left(-1\right)^{x_3}\right)}{2\sqrt{2}} \\
\eta^{(3)}\left(\mathbf{x}\right) &= \frac{1 + i\left(-1\right)^{x_3}}{\sqrt{2}} \\
\label{stageta}
\end{aligned}
\end{equation}
\end{widetext}
which satisfy all the properties from above. 
Therefore it can be used as the  rotation transformation to be used with the Gaussian state we construct. 

\subsubsection{Introducing spin components on-site}

If we insist on a single site translation invariance, we have to use matrices for $\eta^{(i)}$, since phases do not provide what we want, as proven above. 
The rules of composing these $\pi/4$ rotations will force us to make a choice which relies on the $SU(2)$ algebra. 
Given our knowledge from the Dirac theory, the smallest number of spin-components we could choose on site is two, and then, for example, we can fix
\begin{equation}
\eta^{(i)}=\exp\left(i\pi \sigma_i/4\right).
\label{eta2}
\end{equation}

For a physical intuition, we turn again to constructing a Hamiltonian which will be invariant under such transformations. 
This time, we consider a naive discretization of the Dirac Hamiltonian (naive fermions), whose Hamiltonian takes the form
\begin{widetext}
\begin{equation}
H = -\frac{i}{2a}\underset{\mathbf{x},i}{\sum}\Psi^{\dagger}_a\left(\mathbf{x}\right)\left(\alpha_i\right)_{ab}
\left(\Psi_b\left(\mathbf{x}+\hat{\mathbf{e}}_i\right)-\Psi_b\left(\mathbf{x}-\hat{\mathbf{e}}_i\right)\right)
+m \underset{\mathbf{x}}{\sum}\Psi^{\dagger}_a\left(\mathbf{x}\right)\beta_{ab}\Psi_b\left(\mathbf{x}\right)
\label{Halphabeta}
\end{equation}
\end{widetext}
where the spinor $\Psi_a\left(\mathbf{x}\right)$ contains four components, and $\left\{\alpha_i\right\}_{i=1}^3$ and $\beta$ are (Hamiltonian) Dirac matrices. 
We choose them to be 
\begin{equation}
	\alpha_i = \left( {\begin{array}{cc}
			0 & \sigma_i \\
			\sigma_i & 0 \\
	\end{array} } \right), \quad
\beta = \left( {\begin{array}{cc}
		\mathbf{1} & 0 \\
		0 & -\mathbf{1} \\
\end{array} } \right).
\end{equation}

This time, we define the particle hole transformation on the odd sites as
\begin{equation}
\Psi^{\dagger}_a\left(\mathbf{x}\right) \rightarrow 
\left( {\begin{array}{cc}
		0 & \epsilon \\
		\epsilon & 0 \\
\end{array} } \right)_{ab} \Psi_b\left(\mathbf{x}\right),\quad \forall \mathbf{x}:x_1+x_2\quad \text{odd}.
\end{equation}
where $\epsilon =i\sigma_2$. 
The transformed Hamiltonian decouples into two pieces,
\begin{equation}
H=H_{\psi}+H_{\chi}
\end{equation}
describing the dynamics of the two upper components of the transformed spinor $\Psi$ ($\psi$) and the two lower ones ($\chi$), and taking the forms
\begin{widetext}
\begin{equation}
	\begin{aligned}
	H_{\psi}=\left(-\frac{i}{2a}\underset{\mathbf{x},i}{\sum}\psi^{\dagger}_a\left(\mathbf{x}\right)\left(J_i\right)_{ab}
\psi^{\dagger}_b\left(\mathbf{x}+\hat{\mathbf{e}}_i\right) + \text{h.c.}\right) 
	+m \underset{\mathbf{x}}{\sum}\psi^{\dagger}_a\left(\mathbf{x}\right)\psi_a\left(\mathbf{x}\right),\\
		H_{\chi}=\left(-\frac{i}{2a}\underset{\mathbf{x},i}{\sum}\chi^{\dagger}_a\left(\mathbf{x}\right)\left(J_i\right)_{ab}
	\chi_b^{\dagger}\left(\mathbf{x}+\hat{\mathbf{e}}_i\right) + \text{h.c.}\right) 
	-m \underset{\mathbf{x}}{\sum}\chi^{\dagger}_a\left(\mathbf{x}\right)\chi_a\left(\mathbf{x}\right),
	\end{aligned}
\label{Hphichi}
\end{equation}
where
\begin{equation}
	J_i=\sigma_i\epsilon=J^T_i.
\end{equation}
Using the transformation rule with the choice of Eq. \eqref{eta2}, we get that rotation invariance of the Hamiltonians implies that 
\begin{equation}
	\begin{aligned}
		\eta^{(1)T}J^{(1)}\eta^{(1)}=J^{(1)}, \quad \eta^{(1)T}J^{(2)}\eta^{(1)}=J^{(3)}, \quad \eta^{(1)T}J^{(3)}\eta^{(1)}=-J^{(2)}, \\
		\eta^{(2)T}J^{(1)}\eta^{(2)}=-J^{(3)}, \quad \eta^{(2)T}J^{(2)}\eta^{(2)}=J^{(2)}, \quad \eta^{(2)T}J^{(3)}\eta^{(2)}=J^{(1)}, \\
		\eta^{(3)T}J^{(1)}\eta^{(3)}=J^{(2)}, \quad \eta^{(3)T}J^{(2)}\eta^{(3)}=-J^{(1)}, \quad \eta^{(3)T}J^{(3)}\eta^{(3)}=J^{(3)}.
	\end{aligned}
	\label{Jeq}	
\end{equation}
\end{widetext}
Just like any other matrix in the fundamental representation of $SU(2)$,
\begin{equation}
\eta^{(i)T} = \epsilon \eta^{(i)\dagger} \epsilon^T
\end{equation}
and using this along with basic properties of the angular momentum operators, we see that everything works.

We can also choose for $\eta^{(i)}$ higher dimensional representations of $SU(2)$; if we choose them to be four-dimensional matrices we can get the symmetry rules of other doubling prescriptions (e.g. Wilson's fermions~\cite{kogut_lattice_1983}).

The important point of this section, however, is that the rotation rules of fermions in $d=2,3$, when Gaussian fermionic states are constructed, must take spin into account somehow, and even using convenient notions of "spinless" fermions will give rise to the staggering prescription. 
Thus, in any case, these states include a subset which may have a relativistic continuum limit and cannot be treated as "non-relativistic" per-se.

\section{The PEPS construction}\label{sec:peps_construction}

\begin{figure}[t!]
	\centering
	\includegraphics[width=0.7\columnwidth]{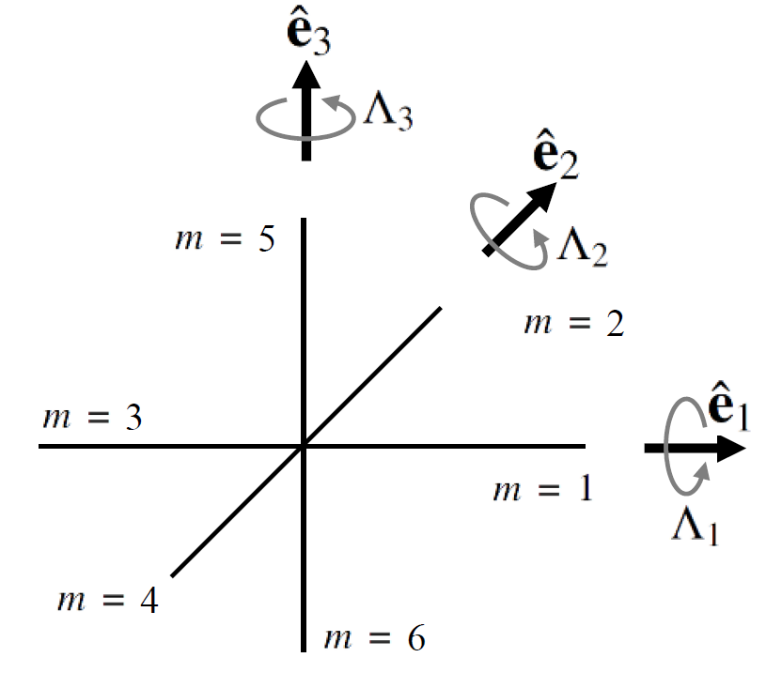}
	\caption{The spatial rotation: The labeling conventions of the legs of a single site ($m$), with respect to the unit vectors $\hat{\mathbf{e}}_i$. When a rotation around $\hat{\mathbf{e}}_i$ is carried out, the coordinate $\mathbf{x}$ is mapped to $\Lambda_i\mathbf{x}$, and the virtual fermionic modes $c$,$d$ associated with the legs are permuted using the matrices $R^{(i)}$ defined in the text (and can be constructed intuitively from this figure).
	}
	\label{figsite}
\end{figure}

After having introduced the symmetries, we are ready to construct the PEPS, following the usual construction procedure of Gaussian fermionic PEPS~\cite{kraus_fermionic_2010}. 
We begin with a "bottom-up" approach, in which we introduce a Gaussian fermionic PEPS construction which takes into account lattice rotation symmetries of the kinds discussed above.

On each site, besides the physical fermions created by $\psi^{\dagger}_a\left(\mathbf{x}\right)$ (again, we assume that the index only labels spin, if it exists, and flavor / spin may be added  as explained in previous works, such as~\cite{zohar_combining_2018}), we introduce auxiliary or \emph{virtual fermionic modes} which will take care of the PEPS contraction. 
As will be explained below, these will be of two types, $c$ and $d$, both associated with all legs going into or out of each site: on each such leg we introduce $\mathcal{N}_s\left(\mathcal{N}_c+\mathcal{N}_d\right)$ auxiliary fermionic modes, where $\mathcal{N}_s$ is the number of spin components and $\mathcal{N}_c,\mathcal{N}_d$ are the numbers of "copies" of $c$ and $d$ fermions on each leg, respectively. 
The latter are integers which control the number of variational parameters we can introduce to our state (this will be multiplied by the number of flavors of colors, in case they are introduced). 
The number of fermionic configurations on each ingoing or outgoing leg, in terms of the virtual fermions, is thus $2^{S\left(\mathcal{N}_c+\mathcal{N}_d\right)}$; in conventional TNs terminology, this is the \emph{bond dimension}~\cite{cirac_matrix_2020}.

In order to introduce the virtual modes quantitatively, let us introduce the index $m=1,...,2d$ to enumerate the  directions going in and out of a site, in the following order: $\hat{\mathbf{e}}_1$, $\hat{\mathbf{e}}_2$, $-\hat{\mathbf{e}}_1$, $-\hat{\mathbf{e}}_2$, $\hat{\mathbf{e}}_3$, $-\hat{\mathbf{e}}_3$. 
We also introduce the index $\mu=1,...,N$ to enumerate the "copies". 
With this at hand, we introduce $c^{(\mu)\dagger}_{ma}\left(\mathbf{x}\right)$ as the $a$ spin component of the $\mu$ labeled copy of the virtual fermionic mode $c$  associated with the $m$ directed leg of site $\mathbf{x}$. 
Similarly, we introduce the $d$ modes, denoted by 
$d^{(\mu)\dagger}_{ma}\left(\mathbf{x}\right)$ (see Fig.~\ref{figsite} for a pictorial representation).

\begin{figure}
    \centering
    \includegraphics[width=\columnwidth]{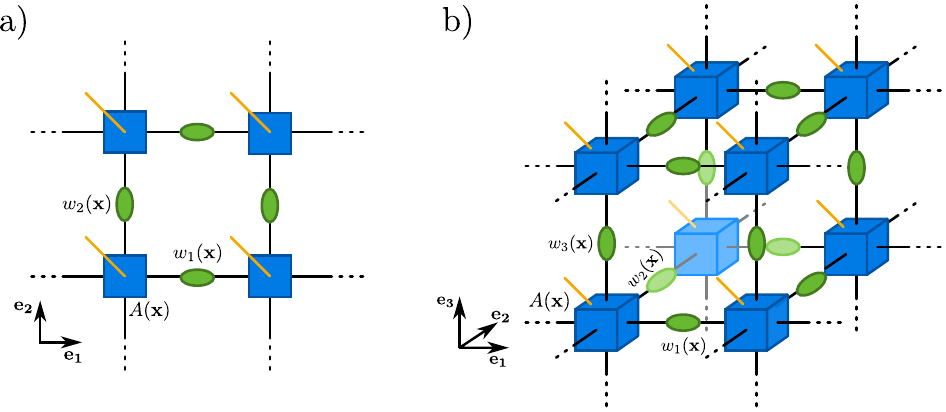}
    \caption{Graphical illustration of the PEPS construction in two space dimensions (a) and three space dimensions (b).
    The operators $A(\vb{x})$ (in blue) act on the sites and the virtual legs are connected via  $w_i(\vb{x})$ (green).
    Virtual legs are drawn in black and physical legs of the PEPS are drawn in yellow.
    The dotted lines illustrate the possibly larger lattice and the periodic boundary conditions.}
    \label{fig:peps}
\end{figure}

On each site, we define the operator
\begin{widetext}
\begin{equation}
A\left(\mathbf{x}\right) = \text{exp}\left(\underset{(\mu)}{\sum}t^{\mu}_{m,ab}\left(\mathbf{x}\right)\psi^{\dagger}_a\left(\mathbf{x}\right)c^{(\mu)\dagger}_{mb}\left(\mathbf{x}\right) + \underset{\mu,\nu}{\sum}\tau^{(\mu,\nu)}_{mn,ab}\left(\mathbf{x}\right)
c^{(\mu)\dagger}_{ma}\left(\mathbf{x}\right)
d^{(\nu)\dagger}_{nb}\left(\mathbf{x}\right)\right)
\label{Adef}
\end{equation}
which defines a Gaussian state involving all the local (physical and virtual) modes when acting on the Fock vacuum $\left|\Omega\right\rangle$, which is the product of the physical and virtual Fock vacua,
\begin{equation}
\left|\Omega\right\rangle=\left|\Omega\right\rangle_{\text{p}}\otimes\left|\Omega\right\rangle_{\text{v}}.
\end{equation}
 Since $\left[A\left(\mathbf{x}\right),A\left(\mathbf{y}\right)\right]=0$ for any two sites $\mathbf{x},\mathbf{y}\in\mathbb{Z}^d$, the product $\underset{\mathbf{x}}{\prod}A\left(\mathbf{x}\right)$ is well defined and requires no ordering.

We further define, on each link $\left(\mathbf{x},i\right)$ connecting the sites $\mathbf{x}$ and $\mathbf{x}+\hat{\mathbf{e}}_i$, the operator
\begin{equation}
w_i\left(\mathbf{x} \right) = \text{exp}
\left(
i
W^{C(i)}_{ab}\left(\mathbf{x}\right)
X^{(i)}_{mn}
\underset{\mu}{\sum}
c^{(\mu)\dagger}_{ma}\left(\mathbf{x}\right)
c^{(\mu)\dagger}_{nb}\left(\mathbf{x}+\hat{\mathbf{e}}_i\right)
\right)
\text{exp}
\left(
i
W^{D(i)}_{ab}\left(\mathbf{x}\right)
X^{(i)}_{mn}
\underset{\mu}{\sum}
d^{(\mu)\dagger}_{ma}\left(\mathbf{x}\right)
d^{(\mu)\dagger}_{nb}\left(\mathbf{x}+\hat{\mathbf{e}}_i\right)
\right)
\label{wdef}
\end{equation} 
\end{widetext}
where $W^{C,D(i)}_{ab}\left(\mathbf{x}\right)$ are three $\mathcal{N}_s\times \mathcal{N}_s$ matrices and $X^{(i)}_{mn}$ are the $2d \times 2d$ matrices,
\begin{equation}
	\begin{aligned}
X^{(1)}_{mn} &=\delta_{m,1}\delta_{n,3}, \\
X^{(2)}_{mn} &=\delta_{m,2}\delta_{n,4}, \\
X^{(3)}_{mn} &=\delta_{m,5}\delta_{n,6}.
\end{aligned}
\end{equation}
Each $w_i\left(\mathbf{x} \right)$ operator, when acting on the virtual Fock vacuum, creates a maximally entangled state involving the virtual modes of both sides of the link. 
These are the states used for the contraction of the PEPS, which is a result of projecting the physical-virtual product state $\underset{\mathbf{x}}{\prod}A\left(\mathbf{x}\right)\ket{\Omega}$ onto the virtual product state
$\underset{\mathbf{x},i}{\prod}w_i\left(\mathbf{x}\right)\ket{\Omega}_{\text{v}}$,
\begin{equation}
\ket{\psi}=\bra{\Omega}_{\text{v}} \underset{\mathbf{x},i}{\prod}w^{\dagger}_i\left(\mathbf{x}\right)
\underset{\mathbf{x}}{\prod}A\left(\mathbf{x}\right)\ket{\Omega}.
\label{PEPSdef}
\end{equation}
This is a fermionic Gaussian state with the form of Eq.~\eqref{psidef}.
The construction of the PEPS is illustrated in Fig.~\ref{fig:peps}.

Since we wish to consider a set of models which include Dirac-like theories, we are interested in Hamiltonians with a global $U(1)$ symmetry, having to do with a conservation of the total number of fermions. 
In the ground state of such models, we are in half filling - the total number of fermions in the system is half of the number of fermionic modes.
In order to obtain the BCS forms we work with, we need to perform some particle-hole transformation on one of the sublattices, e.g. the odd one, as in the examples considered above. 
This changes the global $U(1)$ symmetry, regardless of the spatial dimension and the way we treat spin, to a symmetry under
\begin{equation}
Q=\underset{\mathbf{x}}{\sum}\left(-1\right)^{x_1+...+x_d}\psi^{\dagger}_a\left(\mathbf{x}\right)\psi_a\left(\mathbf{x}\right)
\end{equation}
The half-filling condition becomes, after the particle-hole transformation,
\begin{equation}
	Q\ket{\psi}=0.
\end{equation}
This was already encoded into the PEPS of Eq.~\eqref{psidef}, thanks to the introduction of separate $c,d$ modes, and the way we coupled them in the $A$ and $w_i\left(\mathbf{x}\right)$ operators, defined in Eqs.~(\ref{Adef},\ref{wdef}). 
Let us see why, and explain the need for the separate modes. 
For this, we define the global $U(1)$ transformation
\begin{equation}
	\begin{aligned}
	\psi^{\dagger}_a\left(\mathbf{x}\right) &\rightarrow \text{exp}\left(i\left(-1\right)^{x_1+...+x_d}\theta\right)\psi^{\dagger}_a\left(\mathbf{x}\right),\\
	c^{(\mu)\dagger}_{ma}\left(\mathbf{x}\right) &\rightarrow
	\text{exp}\left(-i\left(-1\right)^{x_1+...+x_d}\theta\right)	c^{(\mu)\dagger}_{ma}\left(\mathbf{x}\right),\\
	d^{(\mu)\dagger}_{ma}\left(\mathbf{x}\right) &\rightarrow
	\text{exp}\left(i\left(-1\right)^{x_1+...+x_d}\theta\right)	d^{(\mu)\dagger}_{ma}\left(\mathbf{x}\right),
	\end{aligned}
\end{equation}
for some $\theta \in \left[0,2\pi\right)$.
Note that the $A\left(\mathbf{x}\right)$ and $w_i\left(\mathbf{x}\right)$ operators, defined in Eqs.~(\ref{Adef},\ref{wdef}), are invariant under this transformation (thanks to the fact that only operators transforming with opposite phases are coupled in the exponentials). 
This immediately gives rise to the desired physical, global $U(1)$ symmetry
\begin{equation}
	e^{i\theta Q}\ket{\psi}=\ket{\psi}.
\end{equation}

We would like to parameterize the PEPS, that is, find general forms of $t,\tau,W^{(i)}$, which will make sure that the state is rotationally invariant - that is,
\begin{equation}
	\mathcal{U}_R \ket{\psi}=\ket{\psi}
	\label{rotinv}
\end{equation}
for any rotation, in both $d=2,3$, using all the options to define the spinors and the rotation transformations discussed above.

\subsection{Rotations in two space dimensions.}
We start with the simpler case, of rotations in two space dimensions. 
Such PEPS were already considered and constructed in Refs.~\cite{zohar_fermionic_2015,zohar_projected_2016}, and here we derive their form in a more rigorous way.

We introduce the unitary $\mathcal{U}_V$ which implements rotations on the virtual fermions, that is
\begin{equation}
	\begin{aligned}
	\mathcal{U}_Vc^{(\mu)\dagger}_{ma}\left(\mathbf{x}\right)\mathcal{U}^{\dagger}_V = \xi_{ab}\left(\mathbf{x}\right)R_{mn}c^{(\mu)\dagger}_{nb}\left(\Lambda\mathbf{x}\right),\\
		\mathcal{U}_Vd^{(\mu)\dagger}_{ma}\left(\mathbf{x}\right)\mathcal{U}^{\dagger}_V = \zeta_{ab}\left(\mathbf{x}\right)R_{mn}d^{(\mu)\dagger}_{nb}\left(\Lambda\mathbf{x}\right).
	\label{rotv}
	\end{aligned}
\end{equation}
Unlike the physical rotation, the virtual one is more complicated and includes not only some spin factors $\xi,\zeta$ analogous to $\eta$, but rather also a spatial ingredient, in the form of the permutation matrix which governs the exchange of the legs under rotations. 
Since we ordered the modes as right, up, left, down, we have
\begin{equation}
R=\left( {\begin{array}{cccc}
		0 & 1 & 0 & 0 \\
		0 & 0 & 1 & 0 \\
		0 & 0 & 0 & 1 \\
		1 & 0 & 0 & 0 \\
\end{array} } \right).
\end{equation}

Using conventional PEPS arguments (see~\cite{zohar_fermionic_2015} for an explanation of the fermionic Gaussian PEPS case), we can guarantee the rotation invariance of the PEPS $\ket{\psi}$ of Eq.~\eqref{PEPSdef}, that is, the fulfillment of Eq.~\eqref{rotinv}, by demanding the invariance of each local operator $A\left(\mathbf{x}\right)$ under rotations,
\begin{equation}
\mathcal{U}_R \mathcal{U}_V A\left(\mathbf{x}\right) \mathcal{U}^{\dagger}_V \mathcal{U}^{\dagger}_R= A\left(\mathbf{x}\right), \quad \forall i,\mathbf{x}
\end{equation}
as well as that of the product of all $w_i\left(\mathbf{x}\right)$ operators, 
\begin{equation}
	 \mathcal{U}_V \underset{\mathbf{x},i}{\prod}w_i\left(\mathbf{x}\right) \mathcal{U}^{\dagger}_V = \underset{\mathbf{x},i}{\prod}w_i\left(\mathbf{x}\right), \quad \quad \forall i
\end{equation}
[the Fock vacua are invariant and thus we get Eq.~\eqref{rotinv}].
Thus, all we need to do is to find $t,\tau$ and $W^{(i)}$ which will satisfy that.

Let us begin with the invariance of the $A\left(\mathbf{x}\right)$ operators. 
We consider, as above, the $d=2$ case with translational invariance, $\mathcal{N}_s=1$ and $\eta^4=-1$.  
Thus $t,\tau$ and $W^{(i)}$ are position independent. 
Under the rotation,
\begin{equation}
	t^{(\mu)}_{m}\psi^{\dagger}\left(\mathbf{x}\right)c^{(\mu)\dagger}_{m}\left(\mathbf{x}\right)
	\rightarrow 
	\eta \xi t^{(\mu)}_{m}\psi^{\dagger}\left(\Lambda\mathbf{x}\right)R_{mn}c^{(\mu)\dagger}_{n}\left(\Lambda\mathbf{x}\right)
\end{equation}
Demanding rotation invariance will give rise to the eigenvalue equation
\begin{equation}
	R^T_{mn}t^{(\mu)}_n = \overline{\eta} \overline{\xi} t^{(\mu)}_m
\end{equation}
(where the coordinates were omitted for simplicity).
There are four eigenvectors of $R$ which we can choose for $t^{(\mu)}_m$. 
Canonical transformations redefining the virtual modes can map one to another, and thus we can choose the simplest option, where all the components of the eigenvector are equal, and the eigenvalue is $1$. 
This implies that
\begin{equation}
	\xi=\overline{\eta}
\end{equation} 
and that $t^{(\mu)}_m = t^{(\mu)}$, independently on $m$.

We move on to parameterize $\tau$. 
Rotation will give rise to 
\begin{equation}
\tau^{(\mu,\nu)}_{mn}
c^{(\mu)\dagger}_{m}
d^{(\nu)\dagger}_{n}
	\rightarrow 
	\overline{\eta}\zeta  \tau^{(\mu,\nu)}_{mn}
	R_{mm'}c^{(\mu)\dagger}_{m'}
	R_{nn'}d^{(\nu)\dagger}_{n'}
\end{equation}
and demanding invariance implies that we need to solve the matrix equation (for every $\mu,\nu$)
\begin{equation}
R^T \tau^{(\mu,\nu)} R = \overline{\zeta}\eta \tau^{(\mu,\nu)}
\end{equation}
which will constrain $\tau$. 
It is very natural to choose $\zeta=\eta$, and then,
for example, if  we choose $\eta=e^{i\pi/4}$ (whose fourth power is $-1$), we obtain 
\begin{equation}
	R^T \tau^{(\mu,\nu)} R =  \tau^{(\mu,\nu)}
\end{equation}
The most general matrices it are, very trivially, the circulant matrices
\begin{equation}
	\tau^{(\mu,\nu)}=\left( {\begin{array}{cccc}
			z_1^{(\mu,\nu)} & z_2^{(\mu,\nu)} & z_3^{(\mu,\nu)} & z_4^{(\mu,\nu)} \\
			z_4^{(\mu,\nu)} & 	z_1^{(\mu,\nu)} & z_2^{(\mu,\nu)} & z_3^{(\mu,\nu)} \\
			 z_3^{(\mu,\nu)} & z_4^{(\mu,\nu)} & z_1^{(\mu,\nu)} & z_2^{(\mu,\nu)} \\
			z_2^{(\mu,\nu)} & z_3^{(\mu,\nu)} & z_4^{(\mu,\nu)}  & z_1^{(\mu,\nu)} \\
	\end{array} } \right),
\end{equation}
for some $z_1^{(\mu,\nu)} , z_2^{(\mu,\nu)} , z_3^{(\mu,\nu)} , z_4^{(\mu,\nu)}\in\mathbb{C}$.

Finally, we have to fix  $W^{C,D(i)}$ which are just numbers in this case. 
One can see that we need to demand that $W^{C(2)}=\overline{\eta}^2 W^{C(1)}$ and thus $W^{C(1)}=1$ and $W^{C(2)}=\overline{\eta}^2$ would make a good choice; similarly, we need $W^{D(2)}=\eta^2 W^{D(1)}$ and thus $W^{C(1)}=1$ and $W^{D(2)}=\eta^2$ would make a good choice.

\subsection{Rotations in three space dimensions.}

The case of three space dimensions is slightly more complicated, but can be handled similarly to the $d=2$ one. 
Here, we want to demand invariance under the three fundamental rotations,
\begin{equation}
	\mathcal{U}^{(i)}_R \ket{\psi}=\ket{\psi}
	\label{rotinv3}
\end{equation}
and introduce, similarly, virtual rotations of the form
\begin{equation}
	\begin{aligned}
	\mathcal{U}_V^{(i)}c^{(\mu)\dagger}_{ma}\left(\mathbf{x}\right)\mathcal{U}^{(i)\dagger}_V &= \xi^{(i)}_{ab}\left(\mathbf{x}\right)R^{(i)}_{mn}c^{(\mu)\dagger}_{nb}\left(\Lambda_i\mathbf{x}\right),\\
\mathcal{U}_V^{(i)}d^{(\mu)\dagger}_{ma}\left(\mathbf{x}\right)\mathcal{U}^{(i)\dagger}_V &= \zeta^{(i)}_{ab}\left(\mathbf{x}\right)R^{(i)}_{mn}d^{(\mu)\dagger}_{nb}\left(\Lambda_i\mathbf{x}\right)	.
	\label{rotv3}
	\end{aligned}
\end{equation}
The "bosonic" part of the rotation now includes the following three permutation matrices (satisfying the right commutation relations)
\begin{equation}
\begin{aligned}
		R^{(1)}&=\left( {\begin{array}{cccccc}
			1 & 0 & 0 & 0 & 0 & 0 \\
			0 & 0 & 0 & 0 & 1 & 0 \\
			0 & 0 & 1 & 0 & 0 & 0 \\
			0 & 0 & 0 & 0 & 0 & 1 \\
			0 & 0 & 0 & 1 & 0 & 0 \\
			0 & 1 & 0 & 0 & 0 & 0 \\
	\end{array} } \right), \\
	R^{(2)}&=\left( {\begin{array}{cccccc}
		0 & 0 & 0 & 0 & 0 & 1 \\
		0 & 1 & 0 & 0 & 0 & 0 \\
		0 & 0 & 0 & 0 & 1 & 0 \\
		0 & 0 & 0 & 1 & 0 & 0 \\
		1 & 0 & 0 & 0 & 0 & 0 \\
		0 & 0 & 1 & 0 & 0 & 0 \\
\end{array} } \right), \\
	R^{(3)}&=\left( {\begin{array}{cccccc}
			0 & 1 & 0 & 0 & 0 & 0 \\
			0 & 0 & 1 & 0 & 0 & 0 \\
			0 & 0 & 0 & 1 & 0 & 0 \\
			1 & 0 & 0 & 0 & 0 & 0 \\
			0 & 0 & 0 & 0 & 1 & 0 \\
			0 & 0 & 0 & 0 & 0 & 1 \\
	\end{array} } \right)
\end{aligned}
\end{equation}
(Fig.~\ref{figsite} may help in understanding how to construct these matrices).
The mode order is chosen as before: $\hat{\mathbf{e}}_2$, $-\hat{\mathbf{e}}_1$, $-\hat{\mathbf{e}}_2$, $\hat{\mathbf{e}}_3$, $-\hat{\mathbf{e}}_3$.
The $\xi^{(i)}_{ab}\left(\mathbf{x}\right),\zeta^{(i)}_{ab}\left(\mathbf{x}\right)$ factors will also have to contribute their share to the non-commutativity of rotations, playing the role of the spin. 
This will depend on the number of spin components we choose.

In any case, following the same steps of $d=2$, 
we can guarantee the rotation invariance of the PEPS $\ket{\psi}$ of Eq.~\eqref{PEPSdef}, that is, the fulfillment of Eq.~\eqref{rotinv}, by demanding the invariance of each local operator $A\left(\mathbf{x}\right)$ under all three rotations,
\begin{equation}
	\mathcal{U}^{(i)}_R \mathcal{U}^{(i)}_V A\left(\mathbf{x}\right) \mathcal{U}^{(i)\dagger}_V \mathcal{U}^{(i)\dagger}_R= A\left(\mathbf{x}\right), \quad \forall i,\mathbf{x}
\end{equation}
as well as that of the product of all $w_i\left(\mathbf{x}\right)$ operators, 
\begin{equation}
	\mathcal{U}^{(j)}_V \underset{\mathbf{x},i}{\prod}w_i\left(\mathbf{x}\right) \mathcal{U}^{(j)\dagger}_V  = \underset{\mathbf{x},i}{\prod}w_i\left(\mathbf{x}\right), \quad \forall i,j
\end{equation}

\subsubsection{The "spinless", staggered case}

We begin with the formulation of staggered fermions, where on each site we have a single physical component and its rotation is made with the staggered phase factors of Eq.~\eqref{stageta}.
Let us start, just like in the $d=2$ case, by 
considering the rotation of the first term in the exponential of $A\left(\mathbf{x}\right)$, coupling the physical and virtual fermions:
\begin{equation}
	\begin{aligned}
	&t^{(\mu)}_{m}\left(\mathbf{x}\right)\psi^{\dagger}\left(\mathbf{x}\right)c^{(\mu)\dagger}_{m}\left(\mathbf{x}\right)
	\rightarrow \\
	&\eta^{(i)}\left(\mathbf{x}\right) \xi^{(i)}\left(\mathbf{x}\right) t^{(\mu)}_{m}\left(\Lambda_i\mathbf{x}\right)\psi^{\dagger}\left(\Lambda_i\mathbf{x}\right)R^{(i)}_{mn}c^{(\mu)\dagger}_{n}\left(\Lambda_i\mathbf{x}\right)
\end{aligned}
\end{equation}
Rotation invariance will give rise to  three ($i=1,2,3$)  equations of the form
\begin{equation}
	R^{(i)T}_{mn}t^{(\mu)}_n\left(\Lambda_i\mathbf{x}\right) =  \overline{\eta^{(i)}\left(\mathbf{x}\right)} \overline{\xi^{(i)}\left(\mathbf{x}\right)}t^{(\mu)}_m\left(\mathbf{x}\right).
\end{equation}
These equations can be solved even for the choice $t^{(\mu)}_m\left(\mathbf{x}\right)=t^{(\mu)}_m$. 
Then we will get three eigenvalue equations: the vector $t^{(\mu)}_m\left(\mathbf{x}\right)=t^{(\mu)}_m$ should be an eigenvector of all three permutation matrices, with eigenvalues 
$\overline{\eta^{(i)}\left(\mathbf{x}\right)} \overline{\xi^{(i)}\left(\mathbf{x}\right)}$. 
The only common eigenvector leads to the solution $t^{(\mu)}_m=t^{(\mu)}$ as in $d=2$, and since the eigenvalue is $1$ we get that
\begin{equation}
\xi^{(i)}\left(\mathbf{x}\right) = \overline{\eta^{(i)}\left(\mathbf{x}\right)}.
\end{equation}

Next, we wish to parameterize $\tau$ - the coupling among the virtual modes. 
Following the case of $t$, we would also like to try and make it position independent. 
The rotation transformation of the relevant part of the $A$ exponential reads
\begin{equation}
	\tau^{(\mu,\nu)}_{mn}
	c^{(\mu)\dagger}_{m}
	d^{(\nu)\dagger}_{n}
	\rightarrow 
	\overline{\eta^{(i)}}\zeta^{(i)}  \tau^{(\mu,\nu)}_{mn}
	R^{(i)}_{mm'}c^{(\mu)\dagger}_{m'}
	R^{(i)}_{nn'}d^{(\nu)\dagger}_{n'}.
\end{equation}
As in the two dimensional case, we set
\begin{equation}
	\zeta^{(i)}\left(\mathbf{x}\right) = \eta^{(i)}\left(\mathbf{x}\right).
\end{equation}
giving rise to the invariance conditions (for each $i$)
\begin{equation}
	R^{(i)T} \tau^{(\mu,\nu)} R^{(i)} =  \tau^{(\mu,\nu)},
\end{equation}
which is solved by the matrices
\begin{equation}
	\tau^{(\mu,\nu)}=\left( {\begin{array}{cccccc}
			z_3^{(\mu,\nu)} & z_1^{(\mu,\nu)} & z_2^{(\mu,\nu)} & z_1^{(\mu,\nu)} & z_1^{(\mu,\nu)} & z_1^{(\mu,\nu)} \\
			z_1^{(\mu,\nu)} & 	z_3^{(\mu,\nu)} & z_1^{(\mu,\nu)} & z_2^{(\mu,\nu)} & z_1^{(\mu,\nu)} & z_1^{(\mu,\nu)} \\
			z_2^{(\mu,\nu)} & z_1^{(\mu,\nu)} & z_3^{(\mu,\nu)} &  z_1^{(\mu,\nu)} & z_1^{(\mu,\nu)} & z_1^{(\mu,\nu)} \\
			z_1^{(\mu,\nu)} & z_2^{(\mu,\nu)} & z_1^{(\mu,\nu)}  & z_3^{(\mu,\nu)} & z_1^{(\mu,\nu)} & z_1^{(\mu,\nu)} \\
			z_1^{(\mu,\nu)} & z_1^{(\mu,\nu)} & z_1^{(\mu,\nu)}  & z_1^{(\mu,\nu)} & z_3^{(\mu,\nu)} & z_2^{(\mu,\nu)} \\
						z_1^{(\mu,\nu)} & z_1^{(\mu,\nu)} & z_1^{(\mu,\nu)}  & z_1^{(\mu,\nu)} & z_2^{(\mu,\nu)} & z_3^{(\mu,\nu)} \\
	\end{array} } \right),
\label{bosrot}
\end{equation}
for some $z_1^{(\mu,\nu)} , z_2^{(\mu,\nu)} , z_3^{(\mu,\nu)} \in\mathbb{C}$.

Finally, we have to fix the values of  $W^{C,D(i)}$. 
This is done from demanding the invariance of $w^{i}\left(\mathbf{x}\right)$ from Eq.~\eqref{wdef} under rotations; however, note that the operators in the exponential of $W^D$ transform like the hopping terms of the creation operators in the particle-hole transformed staggered Hamiltonian of Eq.~\eqref{hsussph}; thus, we can just use (up to proportionality factors) the hopping amplitudes from there here, and fix
\begin{equation}
	W^{D(1)}\left(\mathbf{x}\right)=1,\quad W^{D(2)}\left(\mathbf{x}\right)=i\left(-1\right)^{x_3},\quad W^{D(3)}\left(\mathbf{x}\right)=\left(-1\right)^{x_1+x_2}.
\end{equation}
Since the $c^{\dagger}$ operators transform with $\overline{\eta}$, we can use the hopping amplitudes from the annihilation operators terms of Eq. \eqref{hsussph} for $W^{C}$ and fix
\begin{equation}
	W^{C(1)}\left(\mathbf{x}\right)=1,\quad W^{C(2)}\left(\mathbf{x}\right)=-i\left(-1\right)^{x_3},\quad W^{C(3)}\left(\mathbf{x}\right)=\left(-1\right)^{x_1+x_2}.
\end{equation}

\subsubsection{The \texorpdfstring{$\mathcal{N}_s=2$}{Ns=2} case}

Next, we consider the case of two spin components per fermionic mode. 
$\eta^{(i)}$ are matrices, given by Eq.~\eqref{eta2}, and $\xi^{(i)}$,$\zeta^{(i)}$ will be matrices too. 
We do not assume staggering, so we try and fix all the parameters in a translational invariant way.

The rotation of the physical-virtual fermions coupling term now reads
\begin{equation}
    t^{(\mu)}_{m,ab}\psi_a^{\dagger}\left(\mathbf{x}\right)c^{(\mu)\dagger}_{mb}\left(\mathbf{x}\right)
    \rightarrow 
    \eta^{(i)}_{aa'}\xi^{(i)}_{bb'} t^{(\mu)}_{m,ab}\psi^{\dagger}_{a'}\left(\Lambda_i\mathbf{x}\right)R^{(i)}_{mn}c^{(\mu)\dagger}_{b'n}\left(\Lambda_i\mathbf{x}\right)
\end{equation}
Rotation invariance will give rise to  three ($i=1,2,3$)  equations of the form
\begin{equation}
	R^{(i)}_{mn}t^{(\mu)}_{n,ab} =  \eta^{(i)T}_{aa'}t^{(\mu)}_{m,a'b'} \xi^{(i)}_{b'b}.
\end{equation}
It makes sense to split the space rotations, $m$, and the spin ones $a$. 
Thus we can make the guess
\begin{equation}
t^{(\mu)}_{m,ab} = \hat{t}^{\,(\mu)}_m \, \widetilde{t}^{\ (\mu)}_{ab},
\end{equation}
giving rise to
\begin{equation}
	\begin{aligned}
	&R^{(i)}_{mn} \hat{t}^{\,(\mu)}_n =  \hat{t}^{\,(\mu)}_m  \\
        &  \eta^{\,(i)T}_{aa'}\widetilde{t}^{\ (\mu)}_{a'b'} \xi^{(i)}_{b'b}=\widetilde{t}^{\ (\mu)}_{ab},
	\end{aligned}
\end{equation}
which are solved by $\hat{t}^{\,(\mu)}_m= t^{\mu}$ $\forall m$ and $\widetilde{t}^{\ (\mu)}_{ab} \propto \delta_{ab}$, if we set 
\begin{equation}
	\xi^{(i)}=\overline{\eta^{(i)}}.
\end{equation}
Thus,
\begin{equation}
t^{(\mu)}_{m,ab} = t^{(\mu)}\delta_{ab}, \quad \forall m.
\end{equation}

Next, we wish to parameterize $\tau$ - the coupling among the virtual modes.  
The rotation transformation of the relevant part of the $A$ exponential reads
\begin{equation}
	\tau^{(\mu,\nu)}_{mn,ab}
	c^{(\mu)\dagger}_{ma}
	d^{(\nu)\dagger}_{nb}
	\rightarrow 
	\overline{\eta^{(i)}}_{aa'}\zeta^{(i)}_{bb'}  \tau^{(\mu,\nu)}_{mn,ab}
	R^{(i)}_{mm'}c^{(\mu)\dagger}_{m'a'}
	R^{(i)}_{nn'}d^{(\nu)\dagger}_{n'b'}.
\end{equation}
Following the previous logic, we set
\begin{equation}
	\zeta^{(i)}=\eta^{(i)}
\end{equation}
as well as make the guess
\begin{equation}
	\tau^{(\mu\nu)}_{mn,ab} = \hat{\tau}^{\,(\mu\nu)}_{mn} \, \widetilde{\tau}^{\ (\mu,\nu)}_{ab}.
\end{equation}
This gives rise to the invariance conditions (for each $i$)
\begin{equation}
  \begin{aligned}
    R^{(i)T} \hat{\tau}^{\,(\mu,\nu)} R^{(i)} =  \hat{\tau}^{\,(\mu,\nu)}, \\
    \eta^{(i)^{\dagger}}\widetilde{\tau}^{\ (\mu,\nu)}\eta^{(i)}= \widetilde{\tau}^{\ (\mu,\nu)}
  \end{aligned}
\end{equation}
which are solved by $\hat{\tau}^{\,(\mu,\nu)}$ identical to the $\tau^{(\mu,\nu)}$ of the "spinless" case, given in Eq.~\eqref{bosrot} - which makes sense, since this part is blind to the spin; as well as $\widetilde{\tau}_{ab}^{\ \mu,\nu}=\delta_{ab}$.

Finally, we have to consider the $W^C,W^D$ matrices. 
Using the same logic of the spinless case, this time with the $H_{\phi}$ Hamiltonian of Eq.~\eqref{Hphichi}, we can make the choice
\begin{equation}
\begin{aligned}
	W^{C(i)} &= \overline{J}_i,\\
	W^{D(i)} &= J_i.
\end{aligned}
\end{equation}

One may extend this construction for other types of spinors (e.g. four components which are not naive, Wilson fermions for example, that cannot be broken into their upper and lower components). 
Similar logic and derivations will follow.

The important thing, however, is that we have demonstrated that one can use the formalism of fermionic PEPS to encode states with lattice rotation symmetries which give rise, in a continuum limit (should it exist) to the rotation part of the Lorentz group. 
Such states may be gauged and used as variational ans\"atze for lattice gauge theories, using the methods of~\cite{zohar_combining_2018}.

\section{Exact construction of Dirac Ground States}\label{sec:exact_ground_state}

Finally, to demonstrate the power of the states presented here, we would like to show that they may be useful for constructing \emph{exactly} the ground states of the lattice formulations of Dirac fermions, discussed above, in the Hamiltonians of Eqs.~(\ref{suss2d},\ref{hsussph},\ref{Hphichi}). 
While this top-down construction will involve an infinite number of virtual modes ($\mathcal{N}_c,\mathcal{N}_d \rightarrow \infty$), or an infinite bond dimension in conventional tensor network terms, it does not imply that finite $\mathcal{N}_c,\mathcal{N}_d$ numbers could not approximate them well. 
In general, tensor network states with a sufficiently large bond dimension can approximate any state~\cite{cirac_matrix_2020}, and finite bond dimension approximations will depend on the physics of the approximated state (and its entanglement properties). 
Arguments for the accuracy of approximation and its universality properties for relativistic field PEPS are given in~\cite{shachar_approximating_2022}, but such a discussion is irrelevant here, since the examples given here are of a free state for which this solution is not required, and any analysis of the accuracy of the approximation when $\mathcal{N}_c,\mathcal{N}_d$ is enlarged becomes invalid once gauged and used for studying interacting theories. 
These should be analyzed differently, in a further work.

We begin with a general Hamiltonian which covers all of the above cases, 
\begin{equation}
	\begin{aligned}
	H&=\left(-\frac{i}{2a}\underset{\mathbf{x},i}{\sum}\psi^{\dagger}_a\left(\mathbf{x}\right)\left(K_i\left(\mathbf{x}\right)\right)_{ab}
	\psi^{\dagger}_b\left(\mathbf{x}+\hat{\mathbf{e}}_i\right) + \text{h.c.}\right) \\
	&+m \underset{\mathbf{x}}{\sum}\psi^{\dagger}_a\left(\mathbf{x}\right)\psi_a\left(\mathbf{x}\right),
\end{aligned}	
\end{equation}
where $\psi$ may have a single component or more. In the latter case, $K_i$ will either be a matrix (such as $J_i$), and in the simpler, first case, just a set of numbers, giving rise to the desired Hamiltonian.
In the staggered $d=2$ Hamiltonian of Eq.~\eqref{suss2d}, we have a single spin component per site, $K_1\left(\mathbf{x}\right)=1$ and $K_2\left(\mathbf{x}\right)=i$.
In the staggered $d=3$ Hamiltonian of Eq.~\eqref{hsussph}, we still have a single spin component, and we set
$K_1\left(\mathbf{x}\right)=1$,$K_2\left(\mathbf{x}\right)=i\left(-1\right)^{x_3}$,$K_3\left(\mathbf{x}\right)=\left(-1\right)^{x_1+x_2}$. 
In the naive fermions case we discussed later on, after performing a particle-hole transformation to the $\mathcal{N}_s=4$ Hamiltonian Eq.~\eqref{Halphabeta}, we arrived at the two decoupled $\mathcal{N}_s=2$ Hamiltonians of Eq.~\eqref{Hphichi}. 
The first one, $H_{\psi}$, is obtained when $K_i\left(\mathbf{x}\right)=J_i$. 
Upon performing a simple particle-hole transformation on all the site of modes of the second Hamiltonian, $H_{\chi}$, that is, $\chi \leftrightarrow \chi^{\dagger}$, the same kind of Hamiltonian is obtained, with $K_i\left(\mathbf{x}\right)=\overline{J}_i$.
It can be extended to other relevant cases too, obviously (different numbers of spin components, e.g. $\mathcal{N}_s=4$, and other valid choices of $K_i\left(\mathbf{x}\right)$). 

In all the cases, we assume that the mass is larger than zero, which does not restrict our generality much, since PEPS are mostly useful for gapped states (and only for them if we reduce to the $d=1$ case).
Since the Hamiltonian is gapped, its ground state may be obtained using imaginary time evolution, that is,
\begin{equation}
	\ket{\psi} = e^{-\beta H} \ket{\Omega}_{\text{p}}
	\label{thermal}
\end{equation}
for $\beta\rightarrow\infty$. 
Let $N\rightarrow\infty$ be a very large number, such that 
\begin{equation}
	\epsilon = \frac{\beta}{N}
\end{equation}
satisfies
\begin{equation}
	\epsilon \ll a, m^{-1}.
\end{equation} We split the imaginary time evolution to $N$ steps, across the $N+1$ time steps $t=0,...,N$. 
This defines a spacetime lattice, looking like our usual spatial lattice in the spatial dimensions, copied along the time steps.
 We introduce complex Grassman variables $\theta\left(\mathbf{x},t\right)$ defined on each spacetime lattice site.
 
 Recall the completeness relation of fermionic Grassman variables,
\begin{equation}
	\mathbf{1} = \int d\overline{\theta}d\theta \ket{\theta}\bra{\theta} e^{-\overline{\theta}\theta},
\end{equation}
where $\ket{\theta}$ is a fermionic coherent state, 
\begin{equation}
\ket{\theta} = \left(1+\theta \chi^{\dagger}\right)\ket{\Omega}_a
	\label{cohdef}
\end{equation}
satisfying
\begin{equation}
\chi\ket{\theta} = \theta\ket{\theta}
\end{equation}
for some Grassman $\theta$ and fermionic mode annihilated by $\chi$~\cite{fradkin_field_2013}. 

We denote by $\ket{\Theta_t}$ the multi-mode coherent state of a given $\left\{\theta\left(\mathbf{x},t\right)\right\}$ configuration, for a fixed imaginary time slice $t$, and by $\mathcal{D}\Theta$ an integration over all the $\left\{\theta\left(\mathbf{x},t\right)\right\}$ variables, on all the \emph{spacetime} lattice points.
With this at hand, we can express the ground state from Eq.~\eqref{thermal} in the form
\begin{equation}
	\begin{aligned}
	\ket{\psi} =
	\int \mathcal{D}\Theta \mathcal{D}\overline{\Theta} \ket{\Theta_0}&
	\left[\underset{t=0}{\overset{N-1}{\prod}}\bra{\Theta_t}e^{-\epsilon H} \ket{\Theta_{t+1}}\right]
	\braket{\Theta_N | \Omega_p}\\ \times&
	\exp\left(-\overset{N}{\underset{t=0}{\sum}}\underset{\mathbf{x}}{\sum}\overline{\theta}_a\left(\mathbf{x},t\right)\theta_a\left(\mathbf{x},t\right)\right).
	\end{aligned}
\end{equation}
From here we can proceed in the way commonly used for the derivation of fermionic path integrals using coherent states~\cite{fradkin_field_2013}.
Using the coherent state definition Eq. \eqref{cohdef} we obtain that $\braket{\Theta_N | \Omega_p}=1$.
Let us further evaluate the matrix elements in the middle. 
Since we assume that $\epsilon$ is infinitesimal, we can approximate 
\begin{equation}
	\begin{aligned}
	e^{-\epsilon H} &\approx
	\exp\left(
	\frac{-i\epsilon}{2a}\underset{\mathbf{x},i}{\sum}
	\psi^{\dagger}_a\left(\mathbf{x}+\hat{\mathbf{e}}_i\right)\left(K^T_i\left(\mathbf{x}\right)\right)_{ab}
	\psi^{\dagger}_b\left(\mathbf{x}\right)\right) 
	\\&\times \exp\left(-m\epsilon \underset{\mathbf{x}}{\sum}\psi^{\dagger}_a\left(\mathbf{x}\right)\psi_a\left(\mathbf{x}\right)\right)
	\\&\times \exp\left(
\frac{-i\epsilon}{2a}\underset{\mathbf{x},i}{\sum}
\psi_a\left(\mathbf{x}+\hat{\mathbf{e}}_i\right)
\left(K^{\dagger}_i\left(\mathbf{x}\right)\right)_{ab}
\psi_b\left(\mathbf{x}\right)
\right),
\end{aligned}
\end{equation}
and obtain, using standard properties of fermionic coherent states,
\begin{equation}
	\begin{aligned}
\bra{\Theta_t}e^{-\epsilon H} &\ket{\Theta_{t+1}} = \exp\left(
\frac{-i\epsilon}{2a}\underset{\mathbf{x},i}{\sum}
\overline{\theta}_a\left(\mathbf{x}+\hat{\mathbf{e}}_i,t\right)\left(K^T_i\left(\mathbf{x}\right)\right)_{ab}
\overline{\theta}_b\left(\mathbf{x},t\right)\right) 
\\ &\times\exp\left(r \underset{\mathbf{x}}{\sum}\overline{\theta}_a\left(\mathbf{x},t\right)\theta_a\left(\mathbf{x},t+1\right)\right)
\\ &\times\exp\left(
\frac{-i\epsilon}{2a}\underset{\mathbf{x},i}{\sum}
\theta_a\left(\mathbf{x}+\hat{\mathbf{e}}_i,t+1\right)
\left(K^{\dagger}_i\left(\mathbf{x}\right)\right)_{ab}
\theta_b\left(\mathbf{x},t+1\right)
\right),
\end{aligned}
\end{equation}
where $r=1-m\epsilon$.

Let us consider the $\overline{\theta}\overline{\theta}$ and $\theta\theta$ terms. 
Each of them is defined on a spatial link, at a given time. 
We introduce, on the beginning and end of each such link, Grassman variables $\varphi^{C}_{ma}\left(\mathbf{x},t\right)$ for $t=0,...,N-1$. 
Following our convention, $m=1,2,5$ denote the beginnings of links and $m=3,4,6$ their ends, in the $1,2,3$ directions respectively. 
We also introduce their  complex conjugates $\overline{\varphi}^{C}_{ma}\left(\mathbf{x},t\right)$. 
Similarly, for $t=1,...,N$, we introduce the variables  $\varphi^{D}_{ma}\left(\mathbf{x},t\right)$ and $\overline{\varphi}^{D}_{ma}\left(\mathbf{x},t\right)$. 
These will later be related to the virtual fermionic modes, $c$ and $d$.

Using conventional Gaussian integration techniques~\cite{bravyi_lagrangian_2005}, we can represent our PEPS (as detailed in App.~\ref{appgras}) as
\begin{widetext}
\begin{equation}
	\begin{aligned}
		\ket{\psi}=\int \mathcal{D}\overline{\varphi}\mathcal{D}\varphi  
		&\exp\left(\overset{N-1}{\underset{t=0}{\sum}}\left(-i\underset{\mathbf{x},i}{\sum}X^{(i)}_{nm}\overset{N-1}{\underset{t=0}{\sum}}
		\varphi^C_{ma}\left(\mathbf{x}+\hat{\mathbf{e}}_i,t\right)\left(K^T_i\left(\mathbf{x}\right)\right)_{ab}
		\varphi^C_{nb}\left(\mathbf{x},t\right) -\underset{\mathbf{x}}{\sum}\overline{\varphi^C_{ma}}\left(\mathbf{x},t\right)\varphi^C_{ma}\left(\mathbf{x},t\right)\right) \right) \times \\
		&\exp\left(\overset{N-1}{\underset{t=1}{\sum}}\left(-i\underset{\mathbf{x},i}{\sum}X^{(i)}_{nm}\overset{N}{\underset{t=0}{\sum}}
		\varphi^D_{ma}\left(\mathbf{x}+\hat{\mathbf{e}}_i,t\right)\left(K^{\dagger}_i\left(\mathbf{x}\right)\right)_{ab}
		\varphi^D_{nb}\left(\mathbf{x},t\right) -\underset{\mathbf{x}}{\sum}\overline{\varphi^D_{ma}}\left(\mathbf{x},t\right)\varphi^D_{ma}\left(\mathbf{x},t\right)\right) \right) \times	\\
		&\int \mathcal{D}\overline{\Theta}\mathcal{D}\Theta\exp\left(\underset{\mathbf{x}}{\sum}\left(-\overset{N}{\underset{t=0}{\sum}}\overline{\theta}_a\left(\mathbf{x},t\right)\theta_a\left(\mathbf{x},t\right)+\sqrt{\frac{\epsilon}{2a}}\left[\overset{N-1}{\underset{t=0}{\sum}}\overline{\theta}_a\left(\mathbf{x},t\right)\underset{m}{\sum}\overline{\varphi^C_{ma}}\left(\mathbf{x},t\right)
		+\overset{N}{\underset{t=1}{\sum}}\theta_a\left(\mathbf{x},t\right)\underset{m}{\sum}\overline{\varphi^D_{ma}}\left(\mathbf{x},t\right)\right]\right)\right)
		\ket{\Theta_0}.
	\end{aligned}
	\label{bigeq}
\end{equation}
We can now perform the integration over all the $\theta$ variables in $t>0$ (as shown in App.~\ref{appgras}), and obtain
\begin{equation}
	\begin{aligned}
		&\int \mathcal{D}\overline{\Theta}_0 \mathcal{D}\Theta_0  \ket{\Theta_0}
		\exp\left(-\underset{\mathbf{x}}{\sum}\overline{\theta}_a\left(\mathbf{x},0\right)  \theta_a\left(\mathbf{x},0\right)
		+ \sqrt{\frac{\epsilon}{2a}}\overline{\theta}_a\left(\mathbf{x},0\right)\underset{m}{\sum}\overset{N-1}{\underset{t=0}{\sum}} r^t \overline{\varphi^C_{ma}}\left(\mathbf{x},t\right) + 
		\frac{\epsilon}{2a}\underset{m,n}{\sum}\overset{N-1}{\underset{t=1}{\sum}}\overset{N-1}{\underset{s=t}{\sum}}r^{s-t}\overline{\varphi^D_{ma}}\left(\mathbf{x},s\right)\overline{\varphi^C_{na}}\left(\mathbf{x},t\right)
		\right)	= \\
		&\exp\left(\sqrt{\frac{\epsilon}{2a}}\underset{\mathbf{x}}{\sum}\psi^{\dagger}_a\left(\mathbf{x}\right)\underset{m}{\sum}\overset{N-1}{\underset{t=0}{\sum}} r^t \overline{\varphi^C_{ma}}\left(\mathbf{x},t\right) + 
		\frac{\epsilon}{2a}\underset{m,n}{\sum}\overset{N-1}{\underset{t=1}{\sum}}\overset{N-1}{\underset{s=t}{\sum}}r^{s-t}\overline{\varphi^D_{ma}}\left(\mathbf{x},s\right)\overline{\varphi^C_{na}}\left(\mathbf{x},t\right)\right)\ket{\Omega}_{\text{p}}
	\end{aligned}
\label{bigeq2}
\end{equation}
\end{widetext} 
using the completeness relation of each $\theta_0$ using coherent states, and standard properties of fermionic coherent states.

Next, we introduce virtual fermionic modes, for the PEPS construction, $c^{(\mu)\dagger}_{ma}\left(\mathbf{x}\right)$,$d^{(\mu)\dagger}_{ma}\left(\mathbf{x}\right)$ with $\mathcal{N}_c=N$, $\mathcal{N}_d=N-1$. 
For each $\varphi^C_{ma}\left(\mathbf{x},t\right)$, we define a coherent state $\ket{\varphi^C_{ma}\left(\mathbf{x},t\right)}$ of the mode $c^{(t)\dagger}_{ma}\left(\mathbf{x}\right)$; for each $\varphi^D_{ma}\left(\mathbf{x},t\right)$, we define a coherent state $\ket{\varphi^C_{ma}\left(\mathbf{x},t\right)}$ of the mode $c^{(t)\dagger}_{ma}\left(\mathbf{x}\right)$. 
That is, we associate $\mu$ with the imaginary time. 
We use the completeness relation of all these states, inserted in the middle of Eq.~\eqref{bigeq}, and this can give us immediately an explicit form of the state $\ket{\psi}$ as a PEPS, in the form of Eq.~\eqref{PEPSdef}. 
The parameters of the state are 
\begin{equation}
	t^{(\mu)}=\sqrt{\frac{\epsilon}{2a}}r^{\mu}
\end{equation}
for any dimension and spin representation.
In the "spinless" $d=2$ case we obtain that
\begin{equation}
	z_1^{(\mu,\nu)} = z_2^{(\mu,\nu)} = z_3^{(\mu,\nu)} = z_4^{(\mu,\nu)} = 
	\left\{ {\begin{array}{cc}
			0 & \mu > \nu  \\
			-\frac{\epsilon}{2a}r^{\nu-\mu} & \mu \leq \nu  \\
	\end{array} } \right.
\end{equation}
and similarly, in both the $d=3$ cases we considered,
\begin{equation}
	z_1^{(\mu,\nu)} = z_2^{(\mu,\nu)} = z_3^{(\mu,\nu)} 
	\left\{ {\begin{array}{cc}
			0 & \mu > \nu  \\
			-\frac{\epsilon}{2a}r^{\nu-\mu} & \mu \leq \nu  \\
	\end{array} } \right.
\end{equation}

Finally, considering the $w_i\left(\mathbf{x}\right)$ operators, we can read out by comparing Eq.~\eqref{bigeq} and Eq.~\eqref{wdef}, that in all the cases,
\begin{equation}
	\begin{aligned}
		W^{C(i)}\left(\mathbf{x}\right) &= \overline{K}^{(i)}\left(\mathbf{x}\right),\\
		W^{D(i)}\left(\mathbf{x}\right) &= K^{(i)}\left(\mathbf{x}\right)
	\end{aligned}
\end{equation}
- which is in full accordance with the results we got for the particular cases in the previous section.

While the constructions presented in this section are not unique and should not be seen as the most efficient way to construct the ground states of such Hamiltonians as fermionic Gaussian PEPS, we have shown the existence of such constructions and hence the relevance of such PEPS, when gauged, for the study of lattice gauge theories with fermionic matter.

\section{Summary}
In this work, we saw explicitly why fermionic Gaussian states in a BCS form (which is always valid) must satisfy relativistic lattice rotation properties, in both $d=2$ and $d=3$ space dimensions. 
We showed that such states can be constructed as Gaussian PEPS, in a bottom-up approach, and then, in a top-down approach, showed that this class of states includes, as expected, the exact ground states of several discretizations of the Dirac Hamiltonian.

The results presented here can and will serve as prescriptions, combined with the appropriate gauging mechanism~\cite{zohar_building_2016}, for a variational Monte-Carlo study of ground states of lattice gauge theories using the sign-problem free method presented in~\cite{zohar_combining_2018}.

\begin{acknowledgements}
We would like to thank J. Ignacio Cirac, Norbert Schuch, Tom Shachar, Ariel Kelman and Marco Rigobello for fruitful discussions.
E.Z. acknowledges  the support of the Israel Science Foundation (grant No. 523/20). 
\end{acknowledgements}

\appendix
\section{Grassman Calculus for the exact ground state construction}\label{appgras}
Here we would like to elaborate on  the Grassman Gaussian calculations carried out in section~\ref{sec:exact_ground_state} and complete  missing technical details.

First, note that in order to obtain Eq.~\eqref{bigeq} from the earlier steps, we use conventional Gaussian integration techniques~\cite{bravyi_lagrangian_2005} to re-express the same time, different position exponentials as follows, using the newly introduced fields:
\begin{widetext}
\begin{equation}
	\begin{aligned}
		&\overset{N-1}{\underset{t=0}{\prod}}
		\exp\left(\frac{-i\epsilon}{2a}\underset{\mathbf{x},i}{\sum}
		\overline{\theta}_a\left(\mathbf{x}+\hat{\mathbf{e}}_i,t\right)\left(K^T_i\left(\mathbf{x}\right)\right)_{ab}
		\overline{\theta}_b\left(\mathbf{x},t\right)\right)
		\propto\\ &\int \mathcal{D}\overline{\varphi^C}\mathcal{D}\varphi^C 
		\exp\left(
		\frac{-i\epsilon}{2a}\underset{\mathbf{x},i}{\sum}X^{(i)}_{nm}\overset{N-1}{\underset{t=0}{\sum}}
		\varphi^C_{ma}\left(\mathbf{x}+\hat{\mathbf{e}}_i,t\right)\left(K^T_i\left(\mathbf{x}\right)\right)_{ab}
		\varphi^C_{nb}\left(\mathbf{x},t\right)\right)
		\exp\left(
		-\underset{\mathbf{x}}{\sum}\overset{N-1}{\underset{t=0}{\sum}}
		\left(\overline{\varphi^C_{ma}}\left(\mathbf{x},t\right)\varphi^C_{ma}\left(\mathbf{x},t\right)-\underset{m}{\sum}\overline{\theta}_a\left(\mathbf{x},t\right)\overline{\varphi^C_{ma}}\left(\mathbf{x},t\right)\right)\right).
	\end{aligned}
\end{equation}
and 
\begin{equation}
	\begin{aligned}
		&\overset{N}{\underset{t=1}{\prod}}
		\exp\left(\frac{-i\epsilon}{2a}\underset{\mathbf{x},i}{\sum}
		\theta_a\left(\mathbf{x}+\hat{\mathbf{e}}_i,t\right)\left(K^{\dagger}_i\left(\mathbf{x}\right)\right)_{ab}
		\theta_b\left(\mathbf{x},t\right)\right)
		\propto\\ &\int \mathcal{D}\overline{\varphi^D}\mathcal{D}\varphi^D 
		\exp\left(
		\frac{-i\epsilon}{2a}\underset{\mathbf{x},i}{\sum}X^{(i)}_{nm}\overset{N}{\underset{t=1}{\sum}}
		\varphi^D_{ma}\left(\mathbf{x}+\hat{\mathbf{e}}_i,t\right)\left(K^{\dagger}_i\left(\mathbf{x}\right)\right)_{ab}
		\varphi^D_{nb}\left(\mathbf{x},t\right)\right)
		\exp\left(
		-\underset{\mathbf{x}}{\sum}\overset{N}{\underset{t=1}{\sum}}
		\left(\overline{\varphi^D_{ma}}\left(\mathbf{x},t\right)\varphi^D_{ma}\left(\mathbf{x},t\right)-\underset{m}{\sum}\theta_a\left(\mathbf{x},t\right)\overline{\varphi^D_{ma}}\left(\mathbf{x},t\right)\right)\right)
	\end{aligned}
\end{equation}
where we have neglected proportionality constants, since the state we construct is not normalized anyway.    
In the next step, we rescale the new fields,
\begin{equation}
	\varphi^{C,D} \rightarrow \sqrt{\frac{\epsilon}{2a}}\varphi^{C,D},\quad 
	\overline{\varphi^{C,D}} \rightarrow \sqrt{\frac{2a}{\epsilon}}\overline{\varphi^{C,D}},
\end{equation}
which is allowed since at this point they are independent of one another. 
This brings us to
\begin{equation}
	\begin{aligned}
		&\overset{N-1}{\underset{t=0}{\prod}}
		\exp\left(\frac{-i\epsilon}{2a}\underset{\mathbf{x},i}{\sum}
		\overline{\theta}_a\left(\mathbf{x}+\hat{\mathbf{e}}_i,t\right)\left(K^T_i\left(\mathbf{x}\right)\right)_{ab}
		\overline{\theta}_b\left(\mathbf{x},t\right)\right)
		\propto\\ &\int \mathcal{D}\overline{\varphi^C}\mathcal{D}\varphi^C 
		\exp\left(
		-i\underset{\mathbf{x},i}{\sum}X^{(i)}_{nm}\overset{N-1}{\underset{t=0}{\sum}}
		\varphi^C_{ma}\left(\mathbf{x}+\hat{\mathbf{e}}_i,t\right)\left(K^T_i\left(\mathbf{x}\right)\right)_{ab}
		\varphi^C_{nb}\left(\mathbf{x},t\right)\right)
		\exp\left(
		-\underset{\mathbf{x}}{\sum}\overset{N-1}{\underset{t=0}{\sum}}
		\left(\overline{\varphi^C_{ma}}\left(\mathbf{x},t\right)\varphi^C_{ma}\left(\mathbf{x},t\right)-\sqrt{\frac{\epsilon}{2a}}\underset{m}{\sum}\overline{\theta}_a\left(\mathbf{x},t\right)\overline{\varphi^C_{ma}}\left(\mathbf{x},t\right)\right)\right)
	\end{aligned}
\end{equation}
and                                                                                                                         
\begin{equation}
	\begin{aligned}
		&\overset{N}{\underset{t=1}{\prod}}
		\exp\left(\frac{-i\epsilon}{2a}\underset{\mathbf{x},i}{\sum}
		\theta_a\left(\mathbf{x}+\hat{\mathbf{e}}_i,t\right)\left(K^{\dagger}_i\left(\mathbf{x}\right)\right)_{ab}
		\theta_b\left(\mathbf{x},t\right)\right)
		\propto\\ &\int \mathcal{D}\overline{\varphi^D}\mathcal{D}\varphi^D 
		\exp\left(
		-i\underset{\mathbf{x},i}{\sum}X^{(i)}_{nm}\overset{N}{\underset{t=1}{\sum}}
		\varphi^D_{ma}\left(\mathbf{x}+\hat{\mathbf{e}}_i,t\right)\left(K^{\dagger}_i\left(\mathbf{x}\right)\right)_{ab}
		\varphi^D_{nb}\left(\mathbf{x},t\right)\right)
		\exp\left(
		-\underset{\mathbf{x}}{\sum}\overset{N}{\underset{t=1}{\sum}}
		\left(\overline{\varphi^D_{ma}}\left(\mathbf{x},t\right)\varphi^D_{ma}\left(\mathbf{x},t\right)-\sqrt{\frac{\epsilon}{2a}}\underset{m}{\sum}\theta_a\left(\mathbf{x},t\right)\overline{\varphi^D_{ma}}\left(\mathbf{x},t\right)\right)\right).
	\end{aligned}
\end{equation} 
These lead us directly to Eq.~\eqref{bigeq}.

Next, we perform the integration over $\theta$ variables in $t>0$ (last row of Eq.~\eqref{bigeq}.
As one can easily note, in this row we only have coupling of different times, but not different positions, and thus we can neglect the position coordinate in the following. 
Furthermore, we can define
\begin{equation}
	C_a\left(t\right) = \sqrt{\frac{\epsilon}{2a}}\underset{m}{\sum}\overline{\varphi^C_{ma}}\left(t\right), \quad
	D_a\left(t\right) = \sqrt{\frac{\epsilon}{2a}}\underset{m}{\sum}\overline{\varphi^D_{ma}}\left(t\right).
\end{equation}
We can also drop, for simplicity, all the spin indices for this computation, because all the terms involving $\theta$ are spin-decoupled even for $\mathcal{N}_s>1$. 
With these simplifications, we can start the integration backward in time: we start by integrating the $t=N$ variables, which do not appear in all the exponential terms. 
We have, for each $\mathbf{x}$,
\begin{equation}
	\begin{aligned}
		&\int d\overline{\theta}\left(N\right) d\theta\left(N\right) 
		\exp\left(-\overline{\theta}\left(N\right) \theta\left(N\right) + r \overline{\theta}\left(N-1\right)\theta\left(N\right)+ \theta\left(N\right)D\left(N\right)\right)
	\\	&=\int d\overline{\theta}\left(N\right)\int d\theta\left(N\right)
		\exp\left(-\theta\left(N\right)\left( -\overline{\theta}\left(N\right) + r \overline{\theta}\left(N-1\right) -D\left(N\right)\right)\right)
	\\&	\propto\int d\overline{\theta}\left(N\right) \delta\left(\overline{\theta}\left(N\right),r \overline{\theta}\left(N-1\right) -D\left(N\right)\right) = 1
	\end{aligned}
\end{equation}

For $t=N-1$, we obtain
\begin{equation}
	\begin{aligned}
		&\int d\overline{\theta}\left(N-1\right) d\theta\left(N-1\right) 
		\exp\left(-\overline{\theta}\left(N-1\right) \theta\left(N-1\right) + r \overline{\theta}\left(N-2\right)\theta\left(N-1\right)+ 
		\overline{\theta}\left(N-1\right)C\left(N-1\right)+ 
		\theta\left(N-1\right)D\left(N-1\right)\right)
		\\&=\int d\theta\left(N-1\right)\int d\overline{\theta}\left(N-1\right)
		\exp\left(-\overline{\theta}\left(N-1\right)\left( \theta\left(N-1\right)-C\left(N-1\right)\right)\right) 
		\exp\left(\theta\left(N-1\right)\left(D\left(N-1\right)
		- r \overline{\theta}\left(N-2\right) \right)\right)
		\\&\propto
		\exp\left(
		C\left(N-1\right)
		\left(
		D\left(N-1\right)
		- r \overline{\theta}\left(N-2\right)
		\right)
		\right).
	\end{aligned}
\end{equation}
We move on telescopically, until we get, finally, for each $\mathbf{x}$ an expression which depends only on $\theta\left(0
\right)$ and its conjugate, as well as $C,D$ :
\begin{equation}
	\int d\overline{\theta}\left(0\right) d\theta\left(0\right)  \ket{\theta_0}
	\exp\left(-\overline{\theta}\left(0\right)  \theta\left(0\right)
	+\overline{\theta}\left(0\right)\overset{N-1}{\underset{t=0}{\sum}} r^t C\left(t\right) + 
	\overset{N-1}{\underset{t=1}{\sum}}\overset{N-1}{\underset{s=t}{\sum}}r^{s-t}D\left(s\right)C\left(t\right)
	\right)	,
\end{equation}
and, reintroducing the coordinate, spin and everything we omitted, and identifying $\ket{\theta_0\left(\mathbf{x}\right)}$ as the coherent state of the mode created by $\psi^{\dagger}\left(\mathbf{x}\right)$, we finally obtain Eq.~\eqref{bigeq2}.
\end{widetext}

\bibliography{ref}
\end{document}